\documentclass[aps,pra,amsmath,amssymb,reprint,superscriptaddress]{revtex4-1}

\usepackage[colorlinks,citecolor=blue,linkcolor=red,urlcolor=blue]{hyperref}
\usepackage{graphicx}
\usepackage{subfigure}
\usepackage{verbatim}
\usepackage[normalem]{ulem}
\usepackage{bbm}
\usepackage{physics,amsmath,amsfonts,amssymb,stmaryrd,bm,bbm,bbold}

\usepackage{tikz} \usepackage{physics} \usepackage{stmaryrd}%rrbracket
\usepackage{color}

\preprint{APS/123-QED}

\begin{document}

\title{Effect of top metallic contacts on energy conversion performances \\for near-field thermophotovoltaics} 

\author{Youssef Jeyar}
\email{youssef.jeyar@umontpellier.fr}
\affiliation{Laboratoire Charles Coulomb (L2C), UMR 5221 CNRS-Universit\'{e} de Montpellier, F-34095 Montpellier, France}

\author{Kevin Austry}
\affiliation{Laboratoire Charles Coulomb (L2C), UMR 5221 CNRS-Universit\'{e} de Montpellier, F-34095 Montpellier, France}

\author{Minggang Luo}
\affiliation{Laboratoire Charles Coulomb (L2C), UMR 5221 CNRS-Universit\'{e} de Montpellier, F-34095 Montpellier, France}

\author{Brahim Guizal}
\affiliation{Laboratoire Charles Coulomb (L2C), UMR 5221 CNRS-Universit\'{e} de Montpellier, F-34095 Montpellier, France}

\author{Yi Zheng}
\affiliation{Department of Mechanical and Industrial Engineering, Northeastern University, Boston, MA, 02115, USA}
\affiliation{Department of Chemical Engineering,  Northeastern University,  Boston, MA, 02115, USA}

\author{Riccardo Messina}
\affiliation{Laboratoire Charles Fabry, UMR 8501, Institut d'Optique, CNRS, Universit\'{e} Paris-Saclay, 2 Avenue Augustin Fresnel, 91127 Palaiseau Cedex, France}

\author{Rodolphe Vaillon}
%\affiliation{IES, Univ Montpellier, CNRS, Montpellier, France}
\affiliation{LAAS-CNRS, Universit\'{e}  de Toulouse, CNRS, 7 avenue du Colonel Roche, 31400 Toulouse, France}

\author{Mauro Antezza}
\email{mauro.antezza@umontpellier.fr}
\affiliation{Laboratoire Charles Coulomb (L2C), UMR 5221 CNRS-Universit\'{e} de Montpellier, F-34095 Montpellier, France}
\affiliation{Institut Universitaire de France, 1 rue Descartes, Paris Cedex 05 F-75231, France}

\date{\today}

%------------------------------------------------------------------------------------------------------------------------
\begin{abstract}

The design of metallic contact grids on the front side of thermophotovoltaic cells is critical since it can cause significant optical and electrical resistive losses, particularly in the near field. However, from the theoretical point of view, this effect has been either discarded or studied by means of extremely simplified models like the shadowing methods, that consist in simply ignoring the fraction of the semiconductor surface covered by metal. Our study, based on a rigorous three-body theoretical framework and implemented using the scattering matrix approach with the Fourier modal method augmented with adaptive spatial resolution, provides deeper insight into the influence of the front metal contact grid. This approach allows direct access to the radiative power absorbed by the semiconductor, enabling the proposal of an alternative definition for the thermophotovoltaic cell efficiency. By modeling this grid as a metallic grating, we demonstrate its significant impact on the net radiative power absorbed by the cell and, consequently, on the generated electrical power. Our analysis reveals behaviors differing substantially from those predicted by previous simplistic approaches.

\end{abstract}
%------------------------------------------------------------------------------------------------------------------------
%\pacs{ 02.70.-c, 05.10.-a, 05.45.-a, 64.60.-i }
%\keywords{ , , }

\maketitle
%------------------------------------------------------------------------------------------------------------------------
\section{Introduction}\label{I}
%------------------------------------------------------------------------------------------------------------------------

Radiative heat transfer is the exchange of energy mediated by the electromagnetic field occurring between two bodies at different temperatures and separated by a vacuum gap. Stefan-Boltzmann's law sets an upper limit to this energy flux, achieved only in the ideal scenario of two blackbodies. The advent of fluctuational electrodynamics, dating back to the 1970s and originating from the pioneering works of Rytov~\cite{Rytov53}, Polder and van Hove~\cite{polder1971theory}, showed that this limit does not apply when the distance separating the two bodies is small compared to the thermal wavelength, which is of the order of a few microns at ambient temperature. In this near-field regime, the flux can overcome, by orders of magnitude, the far-field limit (see e.g. Ref.~\cite{joulain2005surface}). This has triggered the idea of exploiting this flux amplification and has been recently investigated for various geometrical configurations, such as nanostructures involving gratings~\cite{PhysRevBNRHT1,PhysRevBNRHT2,APL2023}, for various applications, such as heat-assisted data recording and storage~\cite{Srituravanich04}, infrared sensing and spectroscopy~\cite{DeWilde06,Jones12}, and energy-conversion systems~\cite{DiMatteo01,Narayanaswamy03,Laroche06,Park08} such as thermophotovoltaic (TPV) devices.

The idea of utilizing near-field radiation in TPV systems was introduced in \cite{dimatteo1996enhanced} and \cite{whale1997fluctuational}. Since then, significant progress has been made in both its theoretical understanding and experimental demonstration \cite{song2022modeling,mittapally2023near}.

\begin{center}
\begin{figure}[ht!]
%\hspace*{-0.35cm}
\centering
\begin{tikzpicture}
\draw(0,0) node {\includegraphics[scale=0.28]{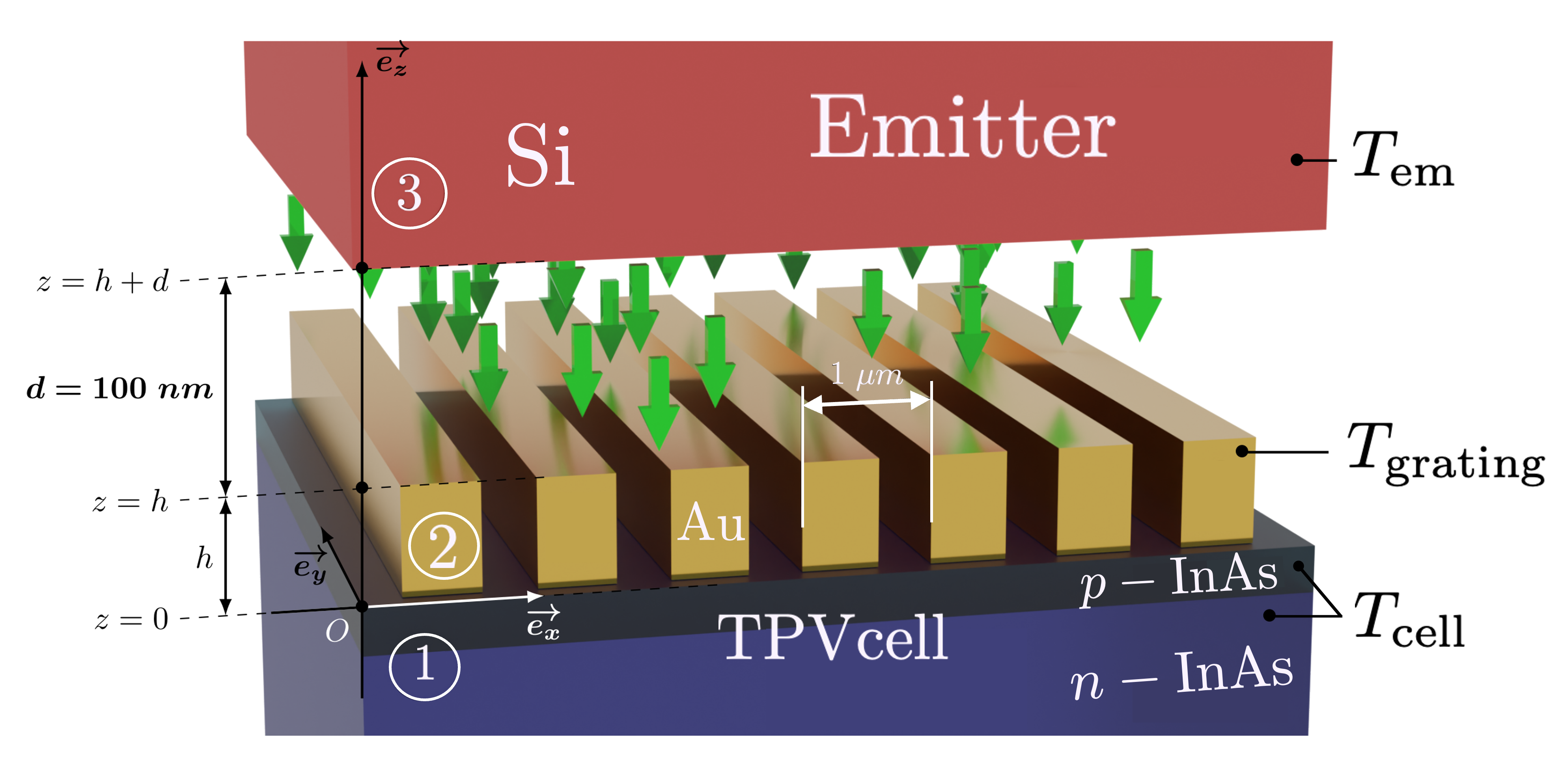}};
\end{tikzpicture}
\caption{Sketch of the geometrical configuration, involving three bodies [InAs semiconductor layers (1), gold grating (2) and p-doped Si emitter (3)] and their characteristic length scales.}\label{geometry}
\end{figure}
\end{center}

In this article, we focus on two main performance metrics: the electrical power generated by the cell, $p_{\rm{out}}$, and the pairwise efficiency, $\eta_{\rm{pairwise}} = {p_{\rm{out}}}/{p_{\rm{rad}}}$, where $p_{\rm{rad}}$ is the net radiative heat flux absorbed by the cell \cite{burger2020present}.

The efficiency of TPV devices operating in the far field has continuously increased, reaching slightly above 40\% with a tandem cell \cite{lapotin2022thermophotovoltaic}, and almost 44\% with a single-junction cell \cite{joule2024efficient}. Despite significant advances achieved through several recent experimental demonstrations \cite{bhatt2020integrated, inoue2021integrated, lucchesi2021near, mittapally2021near}, the efficiency of TPV devices operating in the near field is currently still below 15\%. Among the reasons for this, Joule losses play a critical role \cite{milovich2020design}. The increase in the net radiation flux absorbed by the cell, due to evanescent modes, leads to an increase in the electric current and consequently to unwanted resistive losses at the level of the contacts. In conventional cells, contacts typically include a metallic grid and fingers at the front. In near-field TPV devices, this structure raises issues because the dimensions of the contacts are similar to or even larger than the vacuum gap distance needed to observe substantial near-field enhancements. Additionally, the impact of these front contacts on radiative exchange between the emitter and the cell is usually completely ignored, or at most accounted for by considering the shadowing approximation (i.e., considering the absorption of the semiconductor by its area not covered by metal as in \cite{milovich2020design, song2023effectiveness}).

Using a rigorous approach, the scattering matrix approach (S-matrix) calculated via the Fourier modal method equipped with adaptive spatial resolution (FMM-ASR) \cite{messina2011scattering, messina2014three, GUIZAL1999}, we investigate the real impact of the front metal contact grid on radiative exchange and on conversion performance. By modeling this grid as a metallic grating, we demonstrate its significant effect on the net radiative power absorbed by the active part of the cell. Remarkably, our quantitative and qualitative analyses reveal behaviors substantially different from those predicted by previous simplistic approaches, pointing out the overall inadequacy of the shadowing approximation and the necessity of an accurate theoretical analiys.

The paper is structured as follows: In Section \ref{Physical system}, we describe the physical system. In Section \ref{Modeling}, we introduce our model which is based on a Landauer-like formalism. Finally, in Section \ref{Results and analysis}, we present and discuss the numerical results.

%------------------------------------------------------------------
\section{Physical system}\label{Physical system}
%------------------------------------------------------------------
The physical system, shown in Fig.~\ref{geometry}, consists of three bodies. Body 1 is a p-on-n single-junction TPV cell made of InAs with a bandgap of $0.354$ eV ($\omega_g = 5.38\times 10^{14}$ rad/s), maintained at a temperature of $T_{\rm cell} = 300$ K. The p-doped and n-doped layers of the junction are $400\,$nm and $2.6 \,\mu$m thick, respectively. Body 2 is a gold one-dimensional lamellar grating positioned on top of the cell, characterized by a period $D = 1\,\mu$m, a filling fraction $f$ (i.e., the ratio of the filled to empty part of the grating), and a height $h$. It is maintained at the same temperature ($T_{\rm cell} = T_{\rm grating} = 300$ K). Both structures are invariant along the $y$-axis. Body 3, the emitter, consists of p-doped silicon kept at $T_{\rm em} = 800$ K. It is assumed to be a half-space and is separated from the grating by a fixed distance $d=d_{\rm fixed} = 100$ nm along the $z$-axis. The choice of the emitter and cell materials was made to optimize radiative exchange, as the refractive index of p-doped Si closely matches that of p-doped InAs in the relevant spectral range \cite{milovich2020design}. 
In our study, we vary the height $h$ and the filling fraction of the grating to observe the impact of these parameters on the radiative exchange and the conversion performance of the device.

The dielectric permittivity of gold is calculated using the Drude model $\epsilon(\omega) = 1 - \omega_p^2/(\omega^2 + i\gamma\omega)$ with parameters $\omega_p=137.1 \times 10^{14}$ rad/s and $\gamma=0.405 \times 10^{14}$ rad/s. The optical properties of p-doped and n-doped InAs layers, as well as p-doped silicon, are modeled using parameters and methods from \cite{milovich2020design,ANDERSON1980363,InAs_Otawa}. %An alternative approach is proposed to determine the optical properties of p-doped and n-doped InAs layers \cite{InAs_Otawa}.

The InAs layers emit photons not as blackbodies but as luminescent bodies due to the chemical potential $\mu \approx e\times V$, where $V$ is the voltage of the semiconductor set at $V = 0.9 \dfrac{\hbar \omega_g}{e} \times \qty(1 - \dfrac{T_\text{cell}}{T_\text{em}}) \approx 200$ mV (as in \cite{messina2013graphene}). Each electron-hole pair generated in the semiconductor creates an electron with energy $e\times V$ which we use to define $\hbar \omega_0 = e \times V$. Despite the temperature of the grating and the semiconductor is the same, radiation transfer can still occur between these two bodies due to the non-zero chemical potential of the semiconductor. However, we assume that this radiation transfer is negligible.

%------------------------------------------------------------------
\section{Model}\label{Modeling}
%------------------------------------------------------------------
Starting from Maxwell's equations, we carry out the harmonic representation of the electromagnetic field and perform a plane-wave decomposition into modes, indexed by four parameters $(\omega, \mathbf{k}, p, \phi)$ where $\omega$ is the angular frequency, $p$ the polarization index ($p=1,2$ stands for transversal electric (TE) and transversal magnetic (TM) polarization, respectively), $\phi$ the direction of propagation of the waves $(+,-)$ along the $z$ axis, and $\mathbf{k} = (k_x, k_y)$ the transverse wave vector.

In virtue of the well-known non-additivity of radiative heat transfer, the proper description of energy exchange in our system must be performed within a theoretical framework accounting for many-body effects. In the following, we employ the theory developed in Ref.~\cite{messina2014three}, where each body is represented by a scattering operator describing the reflection $\mathcal{R}$ and transmission $\mathcal{T}$ operators on each side of the body. Assuming that the external environment is thermalized with the cell, the energy received per unit surface and time by bodies $1$ and $2$ are given, respectively, by:
\begin{equation}
\begin{split}
\label{eqn:1}
H_1(d) &= \int_{0}^{\infty}\!\dfrac{\mathrm{d}\omega}{2\pi} \hbar \omega\, n_{31}(\omega) \sum_p \int \dfrac{\mathrm{d}^2\mathbf{k}}{(2\pi)^2} \Tr[\mathcal{T}_p^{31}(\omega, \mathbf{k}, d)],\\
H_2(d) &= \displaystyle \int_{0}^{\infty} \!\dfrac{\mathrm{d}\omega}{2\pi} \hbar \omega\, n_{32}(\omega) \sum_p \int \dfrac{\mathrm{d}^2\mathbf{k}}{(2\pi)^2} \Tr[\mathcal{T}_p^{32}(\omega,  \mathbf{k}, d)],
\end{split}
\end{equation}
where $n_{ij}(\omega) = n_i(\omega, T_i) - n_j(\omega, T_j)$, and $n_i(\omega, T_i) = 1/[\exp(\hbar\omega/k_BT_i) - 1]$. The transmission probability  $\mathcal{T}_p^{31}$ between bodies 3 (emitter) and 1 (semiconductor) and  $\mathcal{T}_p^{32}$ between bodies 3 (emitter) and 2 (grating) can be expressed as \cite{messina2014three}

\begin{widetext}
\begin{equation}
\begin{split}
\mathcal{T}_p^{31}(\omega, k, d)&= U^{(23,1)} \mathcal{T}^{(2)-} U^{(3,2)} \qty(f_{-1}(\mathcal{R}^{(3)-}))U^{(3,2)\dagger} \mathcal{T}^{(2)-\dagger}U^{(23,1)\dagger} \qty(f_{1}( \mathcal{R}^{(1)+}) - \mathcal{T}^{(1)-\dagger} \mathcal{P}_1^{(pw)} \mathcal{T}^{(1)-} ),\\
\mathcal{T}_p^{32}(\omega, k, d) &= f_{-1}(\mathcal{R}^{(3)-}) \Bigl(-U^{(3,2)\dagger} \mathcal{T}^{(2)-\dagger} U^{(23,1)\dagger} f_{1}(\mathcal{R}^{(1)+}) U^{(23,1)} \mathcal{T}^{(2)-} U^{(3,2)}+ U^{(3,12)\dagger} f_{1}(\mathcal{R}^{(12)+}) U^{(3,12)}\Bigr).
\end{split}
\end{equation}
\end{widetext}

Operators $U$ and $f_{-1/1}(\mathcal{R}^{(i)})$, explicitly defined in Ref.~\cite{messina2014three}, are combinations of the reflection and transmission operators of the individual objects $\mathcal{R}^{(i)}$ and $\mathcal{T}^{(i)}$, as well as projectors on the propagative and evanescent sectors of the spectrum. Compared to a system involving only planar interfaces, the presence of the grating opens new channels for radiation transfer. As a consequence of the grating periodicity along the $x$-axis for body 2, a mode is defined by $(\omega, \mathbf{k}_n, p, \phi)$ with $\mathbf{k}_n = (k_{x,n}, k_y)$, where $k_{x,n} = k_x + 2\pi n/D$;  ($n \in \mathbb{Z}$) and $k_x$ takes values in the first Brillouin zone $\qty[-\pi/D, \pi/D]$. The trace appearing in Eq.~\eqref{eqn:1} corresponds to summing the contributions of modes of all orders as follows:
\begin{equation}
\Tr[\mathcal{T}_p(\omega, \mathbf{k}, d)] = \sum_{n=-N}^N\langle \omega, \mathbf{k}_n, p| \mathcal{T}_p|\omega, \mathbf{k}_n, p\rangle.
\end{equation}
The $\mathcal{R}$ and $\mathcal{T}$ operators needed for our calculations are obtained through the FMM-ASR~\cite{Guizal:09, messina2017radiative, PRA_Jeyar}. Within this approach, in the numerical implementations only $2N + 1$ Fourier coefficients are retained, where $N$ is the truncation order. We set this parameter, for the different configurations, to ensure a convergence of the heat fluxes $H_1(N)$ and $H_2(N)$ below $1\%$.

Internal Quantum Efficiency (IQE) of the semiconductor is assumed to be equal to 1. The pairwise conversion efficiency of the TPV device can be determined using two different approaches. Since experiments typically account for the total radiative power absorbed by the entire TPV cell, we define this quantity as the ratio of the output electrical power to the total absorbed radiative power: 
\begin{equation}
\eta = \dfrac{p_\text{out}}{p^{\rm total}_\text{rad}},
\label{eq_eta_total}
\end{equation}
where the total radiative power is expressed as:
\begin{equation}
p^{\rm total}_{\rm rad} = p_{\rm rad}^\text{sc} + p_{\rm rad}^\text{grating}.
\label{eq_p_rad}
\end{equation}

Besides, thanks to the three-body theory we use, \( p^{\rm sc}_\text{rad} \) can be directly accessed. This allows us to propose the following alternative definition for the TPV cell efficiency:

\begin{equation}
\bar{\eta} = \dfrac{p_\text{out}}{p^{\rm sc}_\text{rad}},
\label{eq_eta_bar}
\end{equation}
where 
\begin{equation}
p_{\rm rad}^\text{sc} = p_{\rm rad}^\text{a} -p_{\rm rad}^\text{e}, 
\label{p_rad_sc_eq}
\end{equation}
and the absorbed and the emitted radiative powers are given, respectively, by:
\begin{equation}
\begin{split}
\displaystyle  p_\text{rad}^{a} =&\int_{0}^{\infty} \dfrac{\mathrm{d}\omega}{2\pi} \hbar \omega n(\omega, T_\text{em}) \sum_p \int \dfrac{\mathrm{d}^2\mathbf{k}}{(2\pi)^2} \Tr[\mathcal{T}^{31}_p(\omega, \vb{k}, d)], \\
p_\text{rad}^{e}=&  \int_{0}^{\omega_g} \dfrac{\mathrm{d}\omega}{2\pi} \hbar \omega n(\omega, T_\text{cell}) \sum_p \int \dfrac{\mathrm{d}^2\mathbf{k}}{(2\pi)^2} \Tr[\mathcal{T}^{31}_p(\omega, \vb{k}, d)] \\
+ \int_{\omega_g}^{\infty} &\dfrac{\mathrm{d}\omega}{2\pi} \hbar \omega n(\omega - \omega_0, T_\text{cell}) \sum_p \int \dfrac{\mathrm{d}^2\mathbf{k}}{(2\pi)^2} \Tr[\mathcal{T}^{31}_p(\omega, \vb{k}, d)].
\end{split}
\end{equation}
In the same way, the grating net radiative power is:
\begin{equation}
\begin{split}
\displaystyle p_{\rm{rad}}^\text{grating} & =  \int_{0}^{\infty} \dfrac{\mathrm{d}\omega}{2\pi} \hbar \omega \qty[n(\omega, T_\text{em}) - n(\omega, T_\text{cell})]\\
& \times \sum_p \int \dfrac{\mathrm{d}^2\mathbf{k}}{(2\pi)^2} \Tr[\mathcal{T}_p^{32}(\omega, \vb{k}, d)].
\end{split}
\end{equation}
The electrical power density $p_{\rm{out}}$ can be expressed as the product of the external bias $V$ and the generated photocurrent density $J_\text{ph}$ of the semiconductor, defined as
\begin{equation}
\begin{split}
& p_{\rm{out}} = J_\text{ph} \times V = p_{\rm{out}}^{a} - p_{\rm{out}}^{e}.
\end{split}
\end{equation}
$p_{\text{out}}^{a}$ and $p_{\text{out}}^{e}$ respectively denote the generated electrical and external luminescence powers and are expressed as:

\begin{equation}
\begin{split}
 p_{\rm{out}}^{a}  &= \int_{\omega_g}^{\infty} \dfrac{\mathrm{d}\omega}{2\pi} \hbar \omega_0 n(\omega, T_\text{em}) \sum_p \int \dfrac{\mathrm{d}^2\mathbf{k}}{(2\pi)^2} \Tr[\mathcal{T}^{31}_p(\omega, \vb{k}, d)] \\
p_{\rm{out}}^{e} &=\int_{\omega_g}^{\infty} \dfrac{\mathrm{d}\omega}{2\pi} \hbar \omega_0 n(\omega - \omega_0, T_\text{cell}) \\
&\times \sum_p \int \dfrac{\mathrm{d}^2\mathbf{k}}{(2\pi)^2}\Tr[ \mathcal{T}^{31}_p(\omega, \mathbf{k}, d)].
\end{split}
\end{equation}

One widely used method to consider the presence of metallic contacts on top of the semiconductor is the shadowing approximation \cite{song2023effectiveness, blakers1992shading, saive2018mesoscale}. This approach considers only the radiative and electrical power in regions where the semiconductor is not covered by metal. Consequently, the situation resembles a two-body configuration involving only the silicon emitter and the InAs layers. Thus, within this approximation we can express the power densities and the shadowing efficiency, as
\begin{eqnarray}
p_\text{rad}^\text{shadow} &=& (1 - f) \times p_\text{rad}^\text{2b}, \label{prad_shadow}\\
p_\text{out}^\text{shadow} &=& (1 - f) \times p_\text{out}^\text{2b},\label{pout_shadow}\\
\eta^\text{shadow} &=& \dfrac{p_\text{out}^\text{2b}}{p_\text{rad}^\text{2b}}.
\end{eqnarray}
We remark that, while  $p_\text{rad}^\text{shadow}$ and $p_\text{out}^\text{shadow}$ explicitly depend on the filling fraction $f$, this is not the case for $\eta^\text{shadow}$.
%
%------------------------------------------------------------------
\section{Results and analysis}\label{Results and analysis}
\begin{figure}[ht!]
\centering
\includegraphics[scale=0.35]{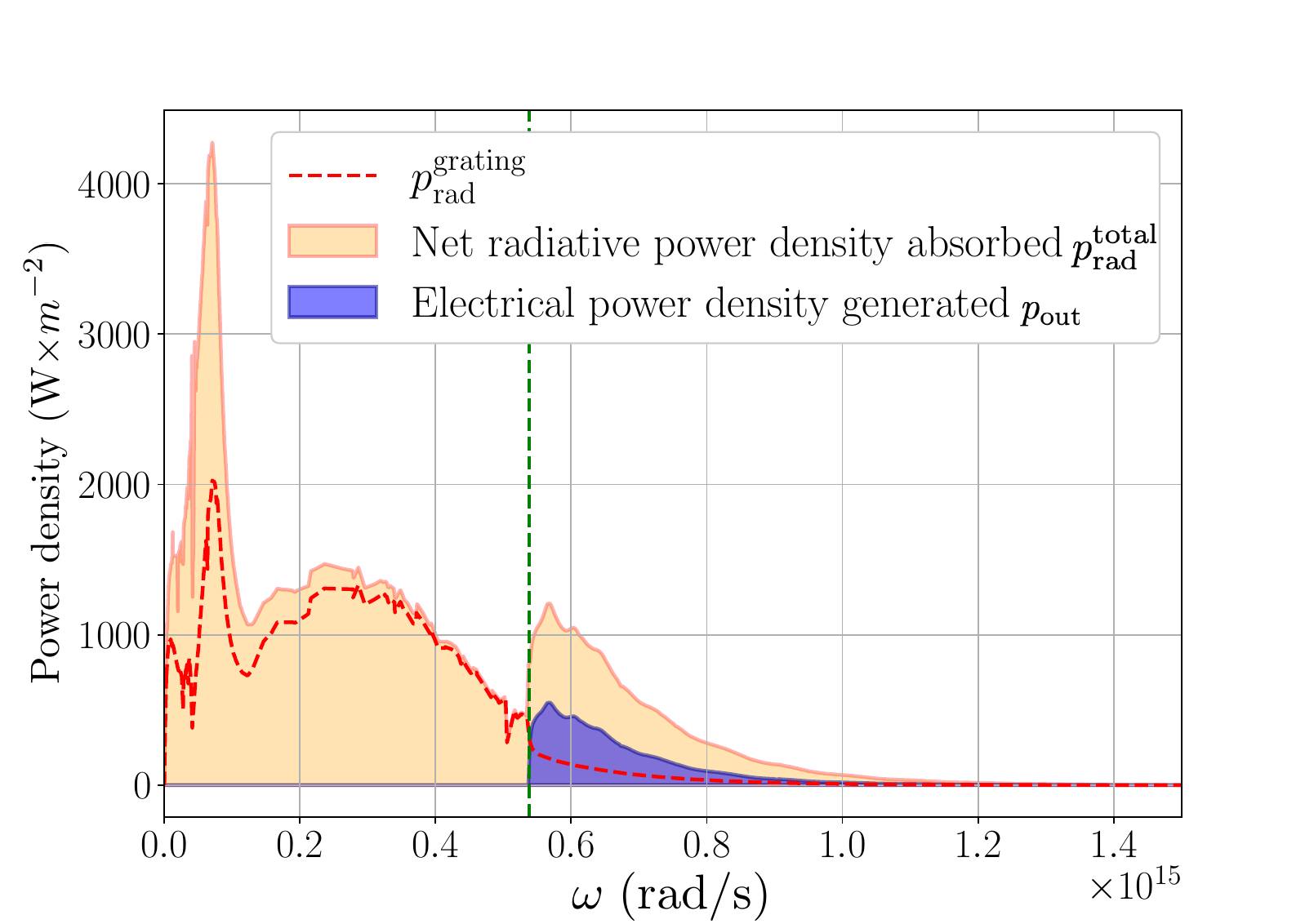}
\caption{Spectrum of the total absorbed radiative power density $p_\text{rad}^{\rm total}$, of the fraction absorbed by the grating alone and of the generated electrical power $p_\text{out}$. The grating has period $D=1\,\mu$m, filling fraction $f=0.5$ and height $h=600\,$nm. The gap thickness between the emitter and the top of the grating is $d=100\,$nm.}
\label{fig:5}
\end{figure}

%------------------------------------------------------------------
Firstly, we emphasize that sub-bandgap photons absorbed by the semiconductor and the grating will reduce the pairwise efficiency because they increase the net radiative power absorbed by the cell, which appears in the denominator of the efficiency calculation. In Fig.~\ref{fig:5}, for a given height $h=600\,$nm of the grating and a filling fraction $f=0.5$, we show the power density spectrum and the distribution between the net absorbed radiative power density $p_\text{rad}$ and the generated electrical power density $p_\text{out}$ on the right. The important point from this figure, and indeed one of the main results in this paper, is the significant radiative power absorbed by the front contacts. This is clearly evidenced by the dashed red curve representing $p_{\rm rad}^{\rm grating}$, which appears in the denominator of the cell's pairwise efficiency $\eta$.

\begin{figure}[ht!]
\centering
 \includegraphics[width=0.5\textwidth]{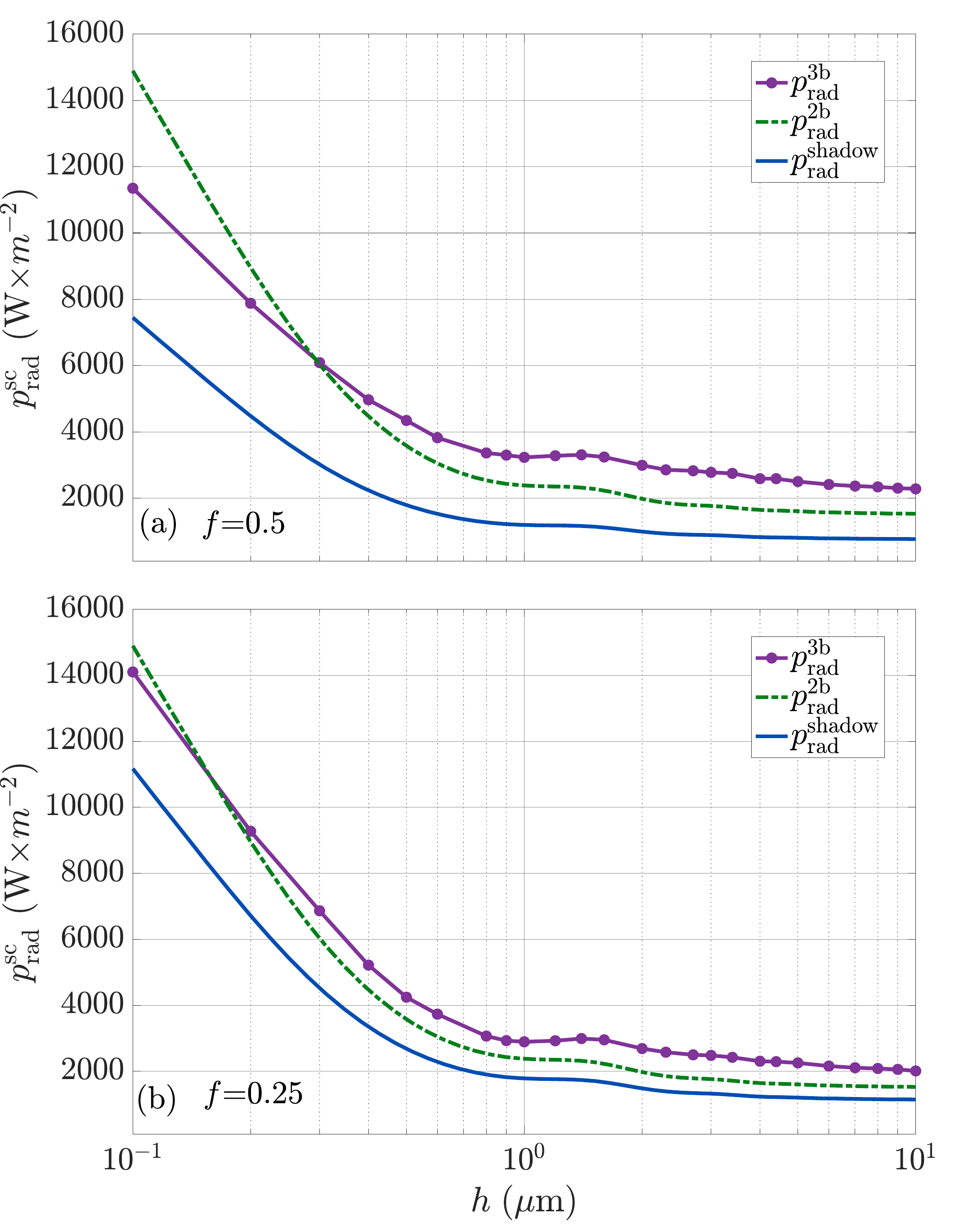}
\caption{Net radiative power density absorbed by the semiconductor as a function of $h$ for the three-body configuration with (a) $f = 0.5$ and (b) $f = 0.25$, $D =\;1 \mu$m, $d_{\rm fixed} = 100$ nm using two different approaches (see main text and legend), and the two-body configuration.}
 \label{fig:p_rad_sc}
\end{figure}

\begin{figure}[ht!]
\centering
 \includegraphics[width=0.48\textwidth]{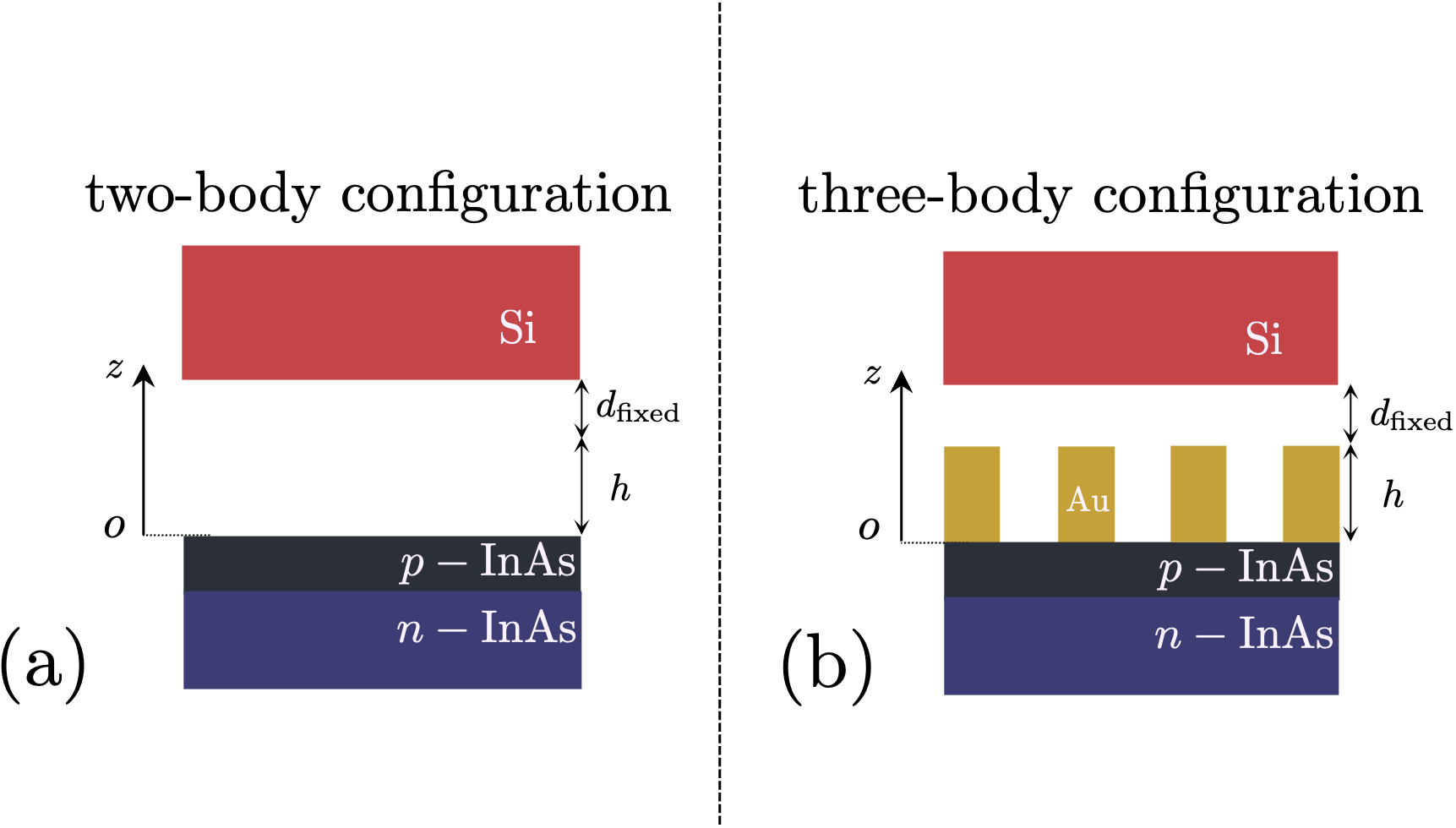}
\caption{The physical system schematics for (a) two-body configuration and (b) three-body configuration.}
 \label{fig:schemas}
\end{figure}

In Fig.~\ref{fig:p_rad_sc}, the net radiative power density absorbed by the semiconductor is shown as a function of $h$ for two different setups: (i) the three-body configuration, which includes the gold grating with $D = 1\;\mu$m, for two filling fractions, $f = 0.5$ (Fig.~\ref{fig:p_rad_sc}(a)) and $f = 0.25$ (Fig.~\ref{fig:p_rad_sc}(b)). Both the exact calculation [Eq. \eqref{p_rad_sc_eq}] and the shadowing approximation [Eq. \eqref{prad_shadow}] are used; and (ii) the two-body configuration without the gold grating. It is worth noticing that, in the three-body configuration, the distance between the emitter and the gold grating is fixed at $d_{\rm fixed} = 100 \, \text{nm}$, and $h$ corresponds to the height of the gold grating. However, in the two-body configuration,  the separation between the emitter and the semiconductor is given by $h + d_{\rm fixed}$. This is illustrated in Fig.~\ref{fig:schemas}(a) and (b).

First, we consider the three-body configuration. The shadowing method (solid blue line) clearly underestimates the net radiative power density absorbed by the semiconductor regardless of the filling fraction. The difference between the exact calculation (violet solid dotted line) and this approximation is more significant when $f = 0.5$, where it can reach approximately 35\% for $h = 100$ nm. This difference is slightly smaller for $f = 0.25$, where it is around 20\% for the same height.

Regardless of the configuration or the filling fraction, the net radiative power density absorbed by the semiconductor decreases with $h$ up to 1\;$\mu$m. Beyond this, a slight increase is observed, reaching a small peak around 1.4 $\mu$m, after which it decreases again with $h$.

The presence of the gold grating enhances the net radiative power density absorbed by the semiconductor when $h$ is greater than 300 nm for $f = 0.5$ and 200 nm for $f = 0.25$, as compared to the two-body configuration (green dashed line). However, this enhancement deteriorates for smaller $h$ values. We explain this behavior by noticing that, in the three-body configuration, the gold grating always remains at a fixed distance of $d_{\rm fixed} = 100$ nm regardless of the values of $h$, ensuring the harvesting of both evanescent and propagative waves. In contrast, in the two-body configuration, the distance between the two bodies is $h + d_{\rm fixed}$. As $h$ increases, the overall thickness of the vacuum gap increases. Thus, the influence of evanescent waves decreases, and only the propagative waves contribute to the net radiative power density absorbed by the semiconductor.

\begin{figure}[ht!]
\centering
 \includegraphics[width=0.5\textwidth]{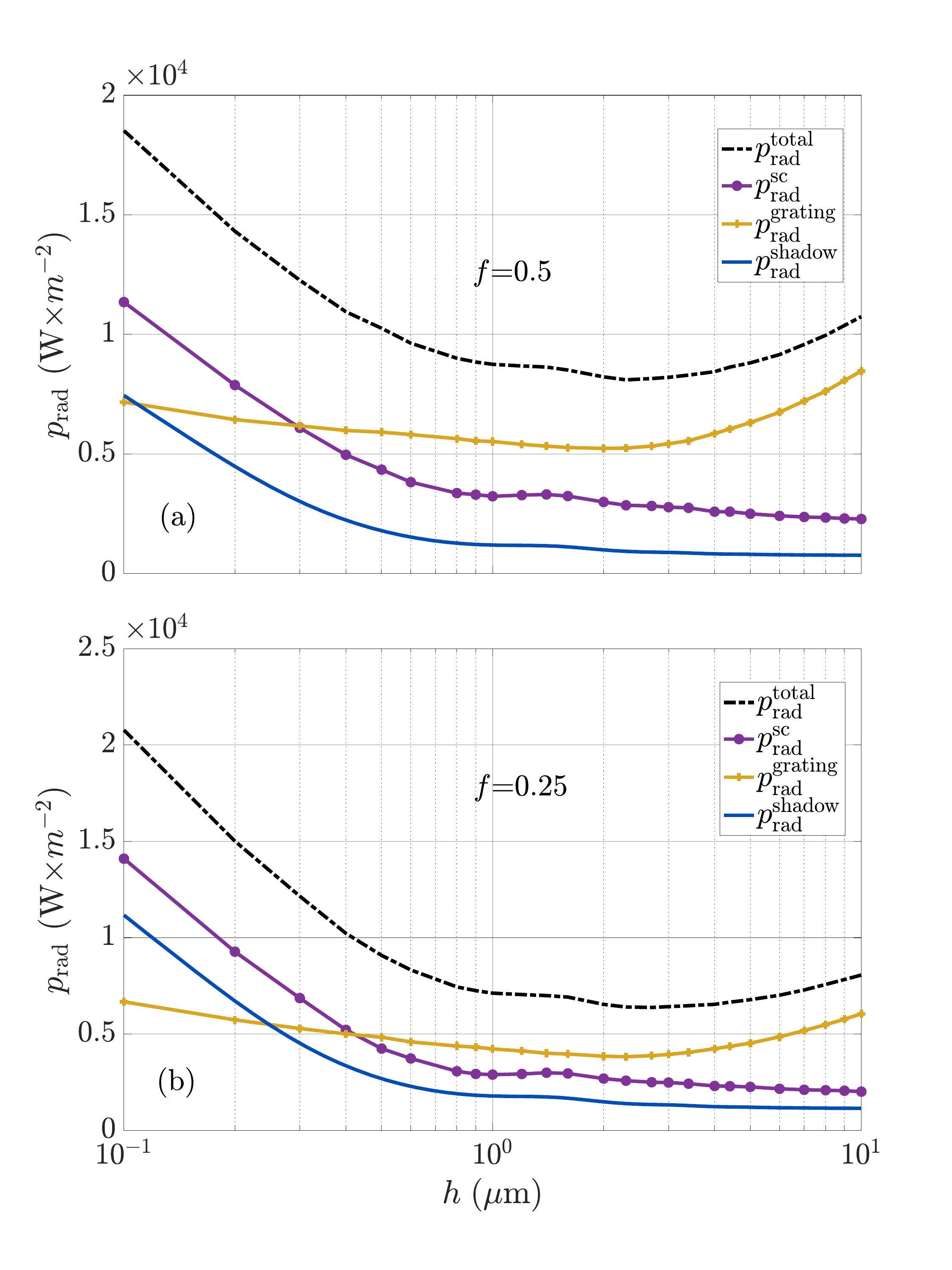}
 \caption{Net radiative power densities calculated for the filling fraction (a) $f = 0.5$ and (b) $f = 0.25$ as a function of $h$ for the three-body configuration. The radiative power density absorbed by the semiconductor and by the grating are plotted for comparison, using the same parameters as in Fig.~\ref{fig:p_rad_sc}.}
 \label{fig:PV_rad_tot}
\end{figure}

Next, we discuss the net radiative power density absorbed by the grating, shown (with the gold solid line with pluses markers) in Fig.~\ref{fig:PV_rad_tot}(a) and Fig.~\ref{fig:PV_rad_tot}(b) for two filling fractions, $f = 0.5$ and $f = 0.25$, respectively. It can be seen that the radiative power density absorbed by the grating for $f = 0.5$ is higher than that for $f = 0.25$, regardless of $h$, due to the larger fraction of gold.

\begin{figure}
 \centering
 \includegraphics[width=0.5\textwidth]{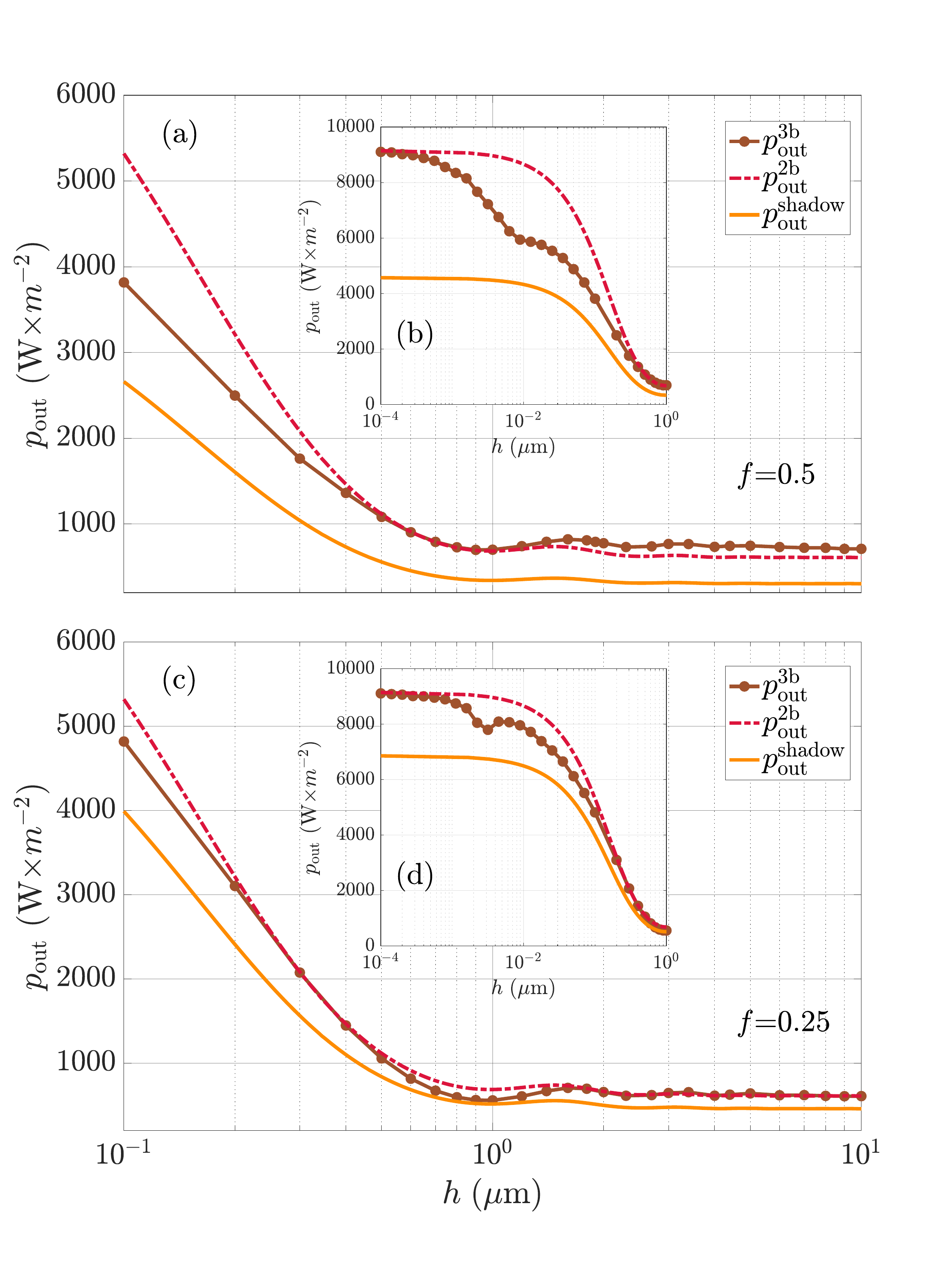}
 \caption{Electrical power density generated by the TPV cell as a function of $h$ for the three-body configuration with (a) $f = 0.5$ and (c) $f = 0.25$, $D = 1 \;\mu$m, $d_{\rm fixed} = 100$ nm using two different approaches (see main text and legend), and the two-body configuration. Insets (b) and (d) show the electrical power generated by the TPV cell for $h$ tending towards 0.}
 \label{fig:PVelectrical}
\end{figure}

In the same figure, we plot the net total radiative power density absorbed by the TPV cell (black dashed line), and, for comparison, the net radiative power density absorbed by the semiconductor. For $f = 0.5$, the contribution of the net radiative power density absorbed by the semiconductor is higher than that of the gold grating for $h$ values below 300 nm; beyond this, the gold grating contribution dominates. For $f = 0.25$, this transition occurs at $h$ values below 400 nm. Therefore, for $f = 0.5$, the total radiative power density is lower than that for $f = 0.25$ when $h$ is below 300 nm; beyond 300 nm, it becomes higher.

Regardless of the filling fraction and $h$, the shadowing technique underestimates the net total power density absorbed by the cell, as compared to the exact calculation. For example, the difference compared to the exact calculation is about 75\% for $h = 300$ nm and 90\% for $h = 10 \;\mu$m if $f = 0.5$, and around 60\% and 80\% for $h = 300$ nm and $h = 10 \;\mu$m, respectively, if $f = 0.25$.

Let us now study the electrical power generated by the TPV cell for the three-body configuration using both the exact calculation and the shadowing approximation, as well as the two-body configuration. The electrical power generated by the TPV cell is shown in Fig.~\ref{fig:PVelectrical}. For $f = 0.5$ (Fig.~\ref{fig:PVelectrical}(a)), the shadowing technique underestimates the electrical power generated by the TPV cell, regardless of $h$. For example, this difference is about 30\% for $h=100$ nm and about 50\% for $h=1 \mu$m. This difference decreases for $f = 0.25$ (Fig.~\ref{fig:PVelectrical}(c)), where it is about 17\% for $h=100$ nm, and for $h$ between 800 nm and 1 $\mu$m becomes negligible.

In the insets of Fig.~\ref{fig:PVelectrical}(a) and Fig.~\ref{fig:PVelectrical}(b), we report the electrical power generated by the TPV cell for $h$ tending towards 0. It can be clearly seen that the three-body configuration tends towards the two-body configuration regardless of the filling fraction, which confirms our exact calculation as opposed to the calculation using the shadowing technique.

\begin{figure}[ht!]
\centering
 \includegraphics[width=0.45\textwidth]{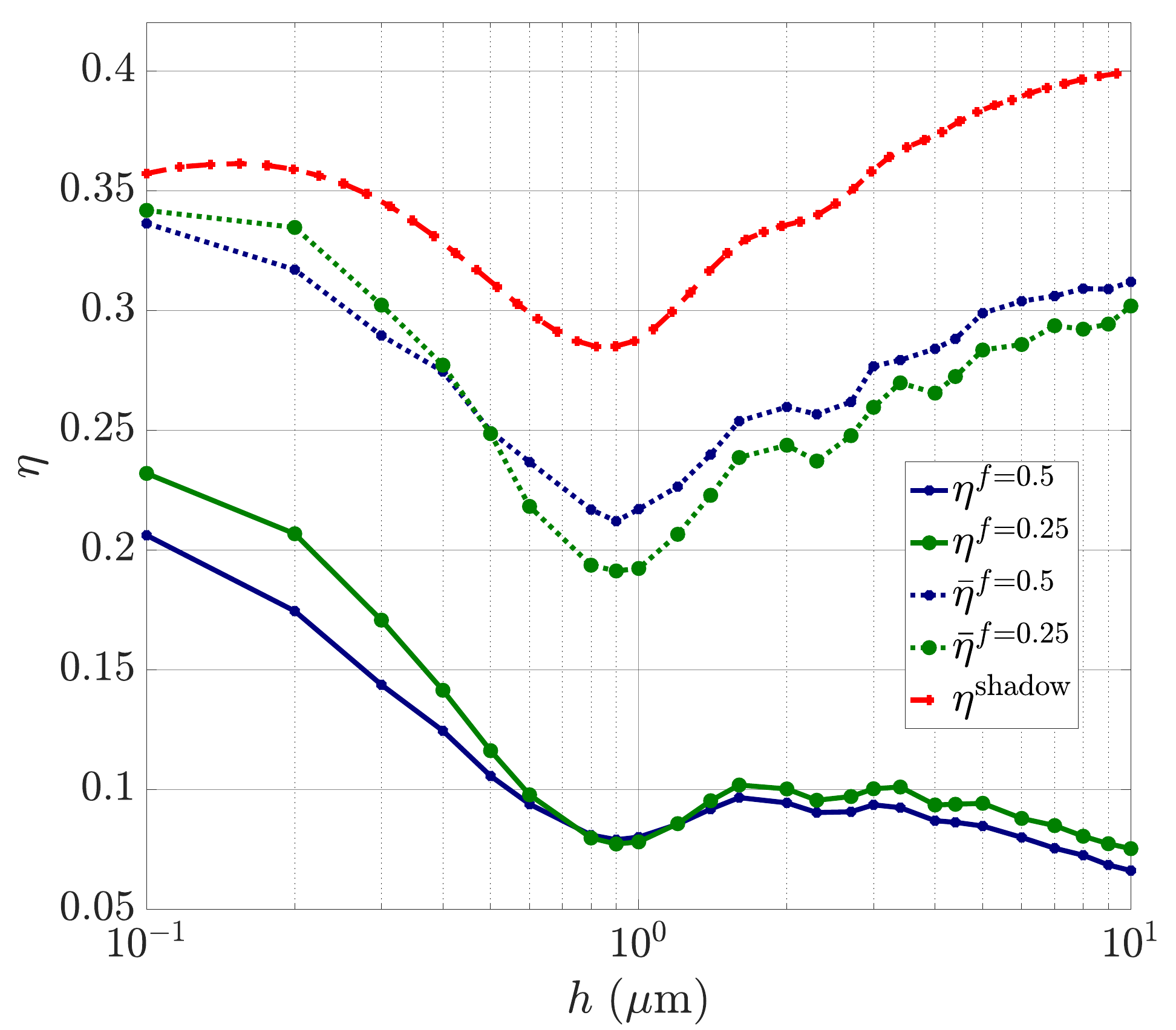}
 \caption{Efficiency $\eta$ [calculated with Eq. \eqref{eq_eta_total}] and $\bar{\eta}$ [calculated with Eq. \eqref{eq_eta_bar}] of the TPV cell for two filling fractions, $f = 0.5$ and $f = 0.25$, compared with the shadowing technique, using the same parameters as in Fig.~\ref{fig:p_rad_sc}.}
 \label{fig:eta}
\end{figure}

Finally, we analyze the efficiency of the TPV cell as a function of $h$ for different filling fractions and approaches, as shown in Fig.~\ref{fig:eta}. The difference between $\eta$ accounting for the total radiative power absorbed by both the semiconductor and the gold grating and the one stemming from the shadowing approximation $\eta^{\rm shadow}$ is significant. The shadowing technique systematically overestimates the efficiency regardless of $h$, with a difference of around 75\% at $h = 100$~nm and 400\% at $h = 10\,\mu$m for a filling fraction of $f = 0.5$.  

When comparing $\bar{\eta}$ (which depends only from the radiative power absorbed by the semiconductor) with the shadowing approximation, the difference is also present but less pronounced, especially for $h$ below 300~nm. For instance, the difference is around 5\% for $h = 100$~nm, 45\% at $h = 1\,\mu$m, and approximately 30\% at $h = 10\,\mu$m for $f = 0.5$. Notably, the difference between $\bar{\eta}$ and $\eta^{\rm shadow}$ is less significant, and both efficiencies exhibit similar trends. This is expected as, in both cases, the radiative power absorbed by the gold grating is not accounted for.  

Since it is $\eta$ that can be experimentally measured, we focus on its analysis for the two filling fractions. For $h < 800$~nm, the efficiency is higher for $f = 0.25$ compared to $f = 0.5$ due to reduced absorption by gold. This highlights that lowering the grating height can enhance the cell's efficiency. Between 800~nm and 1.4~$\mu$m, $\eta$ becomes independent of $f$, while for $h > 1.4\,\mu$m, the efficiency for $f = 0.25$ again surpasses that for $f = 0.5$.   
%------------------------------------------------------------------
\section{Conclusion}
%------------------------------------------------------------------
We have studied the near-field radiative transfer in a TPV device where the metallic front contacts of the cell are modeled by a periodic 1D grating. The uniqueness of this study lies in considering the radiative flux absorbed by the InAs layers and the electrode upon it in determining the cell performances. By employing a rigorous three-body theoretical framework, we have shown that this approach allows direct access to the radiative power absorbed specifically by the semiconductor, providing an accurate description of the energy transfer mechanisms within the TPV cell. This level of detail has enabled us to propose an alternative definition of efficiency that isolates the contribution of the semiconductor. Furthermore, we have shown that the shadowing technique is not sufficient to describe this phenomenon, as it fails to account for key aspects of the radiative interactions in the system. The height $h$ of the grating plays a crucial role in radiative transfer and the electrical power density generated by the semiconductor. Ideally, it should be less than $500\ \rm{nm}$, and the filling fraction of the grating should also be low, even though there is a compromise to consider in the transport of electric charges through the metal. 
Possible extensions of this investigation can be to use different semiconductor voltages to determine the current-voltage characteristic (I-V curve) and find the voltage at which the generated electrical power density is maximum. One can also add, in addition to the radiative recombination included in our analysis, other recombination mechanisms, like the Auger recombination, the Shockley-Read-Hall recombination, the surface recombination and additional series resistance caused by the substrate, lateral transport in the p-doping layer and the metal grating. Finally, taking into account a cell back mirror would allow for the recycling of photons back to the emitter or the creation of other electron-hole pairs in the semiconductor layers and thus further increasing the conversion efficiency of the device. 

\section{ACKNOWLEDGMENTS}
We acknowledge Mathieux Giroux, Jacob Krich and Raphael St-Gelais for fruitful discussions.
Y.J., M.L., B.G., and M.A. acknowledge the support by a grant "CAT" from the ANR/RGC Joint Research Scheme sponsored by the French National Research Agency (ANR) and the Research Grants Council (RGC) of the Hong Kong Special Administrative Region, China (Project No. A-HKUST604/20).

\bibliography{Article_v8.bib}

\end{document}